\documentclass[12pt]{article}

\catcode`\@=11
\@addtoreset{equation}{section}

\global\arraycolsep=2pt

\usepackage{jheppub,amsbsy,amssymb,latexsym,amsfonts,amsmath,epsfig,multicol,bbm}

\allowdisplaybreaks

\newcommand{\al}{\alpha}
\newcommand{\be}{\beta}
\newcommand{\ga}{\gamma}
\newcommand{\de}{\delta}
\newcommand{\ep}{\epsilon}

\newcommand{\Tr}{{\rm Tr}}
\newcommand{\str}{{\rm STr}}

\newcommand{\alg}[1]{\mathfrak{#1}}
\newcommand{\el}{\nonumber}

\newcommand\equ[1] {\begin{equation}#1\end{equation}}

\renewcommand\) {\right)}

\title{Deformations of $T^{1,1}$ as Yang-Baxter sigma models} 

\author{P. Marcos Crichigno$^{\dagger}$\footnote{E-mail:~p.m.crichigno@uu.nl},   
Takuya Matsumoto$^{\dagger}$\footnote{E-mail:~t.matsumoto@uu.nl},  
and Kentaroh Yoshida$^{\ast}$\footnote{E-mail:~kyoshida@gauge.scphys.kyoto-u.ac.jp}}
\affiliation{
$^{\dagger}${\it Institute for Theoretical Physics and Spinoza Institute, 
Utrecht University, \\ 
Leuvenlaan 4, 3854 CE Utrecht, The Netherlands.} 
\vspace*{0.25cm}\\ 
$^{\ast}${\it Department of Physics, Kyoto University, \\ 
Kyoto 606-8502, Japan.} 
}

\arxivnumber{1406.2249}

\abstract{
We consider a family of deformations of $T^{1,1}$ in the Yang-Baxter sigma model approach. 
We first discuss a supercoset description of $T^{1,1}$, which makes manifest the full symmetry 
of the space and leads to the standard Sasaki-Einstein metric.
Next, we consider three-parameter deformations of $T^{1,1}$ 
by using classical $r$-matrices satisfying the classical Yang-Baxter equation (CYBE). 
The resulting metric and NS-NS two-form agree exactly with the ones obtained via TsT transformations, 
and contain the Lunin-Maldacena background as a special case. 
It is worth noting that for AdS$_5\times T^{1,1}$\,, 
classical integrability for the full sector has been argued to be lost.  
Hence our result indicates that the Yang-Baxter sigma model approach is applicable even for  
non-integrable cosets. This observation suggests that the gravity/CYBE correspondence 
can be  extended beyond integrable cases.     
}


\begin{document}

\maketitle 

\section{Introduction}

A fascinating subject in string theory is dualities between gravitational theories and gauge theories. 
The original form proposed in \cite{M} is the AdS/CFT correspondence, stating a duality 
between type IIB string theory on AdS$_5\times$S$^5$ and $\mathcal{N}=4$ $SU(N)$ 
super Yang-Mills (SYM) theory in four dimensions. The integrable structure behind AdS/CFT 
plays a significant role in this duality \cite{review}. It enables one to exactly compute some physical quantities 
such as anomalous dimensions and scattering amplitudes, even at finite coupling without supersymmetries. 

\medskip 

Here we are concerned with the string theory side of the correspondence. 
In the Green-Schwarz formalism, the classical action for the AdS$_5\times$S$^5$ superstring 
is given by a 2d $\sigma$-model on the coset superspace \cite{MT},
\begin{eqnarray}
 \frac{PSU(2,2|4)}{SO(1,4)\times SO(5)}\,. 
\end{eqnarray} 
Classical integrability for the AdS$_5\times$S$^5$ superstring is closely related to 
the existence of a $\mathbb{Z}_4$-grading \cite{BPR}. For an argument of integrability
based on the Roiban-Siegel formalism \cite{RS}, see \cite{Hatsuda}. 
A classification of possible integrable cosets is given in \cite{Zarembo-symmetric}. 

\medskip 

Recently, there has been progress in the study of  integrable deformations 
of the AdS$_5\times$S$^5$ superstring. The Yang-Baxter sigma model approach \cite{Klimcik} 
(generalized to the coset case in \cite{DMV}) plays an important role in this direction. 

\medskip 
 
A $q$-deformed action for the AdS$_5\times$S$^5$ superstring has been 
constructed in \cite{DMV2}. Since a bosonic subsector of this action exhibits 
a $q$-deformed $\mathfrak{su}(2)$, the full symmetry algebra is expected 
to be a $q$-deformed $\mathfrak{psu}(2,2|4)$ \cite{DMV,KYhybrid}
\footnote{
It would be nice to show an affine extension of $\mathfrak{psu}(2,2|4)$ 
by following the procedure \cite{KMY-QAA,KOY}.}. 
In the end, the deformation used in \cite{DMV2} is the standard one with the classical $r$-matrix of 
Drinfeld-Jimbo type \cite{Drinfeld1,Jimbo}. The metric in the string frame and NS-NS two-form 
were obtained in \cite{ABF}, though the complete supergravity solutions have not been found yet. 
Some limits of the deformed background are considered in \cite{HRT,AvT}. 
A mirror TBA is discussed in \cite{AdLvT}. 
A non-relativistic limit on the world-sheet is considered in \cite{Kame}. 
Notably, the singularity of the metric disappears in this limit. 
Giant magnons are constructed in \cite{AdLvT,Magnon}.   

\medskip 

One may consider non-standard $q$-deformations 
(often called Jordanian deformations) \cite{Jordanian,KLM} as well. 
Jordanian-deformed actions for AdS$_5\times$S$^5$ have been constructed in \cite{KMY-Jor}. 
The deformations are characterized by classical $r$-matrices satisfying 
the classical Yang-Baxter equation (CYBE). 
So far, some $r$-matrices, corresponding to well-known string backgrounds 
such as Lunin-Maldacena-Frolov backgrounds \cite{LM,Frolov}, 
and the gravity duals of noncommutative gauge theories \cite{HI,MR}, 
have been found in \cite{MY-LM} and \cite{MY-MR}, respectively\footnote{
The fermionic sector has not been studied yet, simply due to some technical complications. 
To do so, one would have to perform a supercoset construction in the supermatrix notation 
to evaluate the $R$-operator. It would be an important task to complete the analysis.}. 
A new gravitational solution\footnote{It contains 3D Schr\"odinger spacetime. 
The related integrable structure is studied in \cite{KY-Sch}.} 
was also constructed from an $r$-matrix in \cite{SUGRA-KMY}. 
The relation between gravitational solutions and classical $r$-matrices 
may be referred to as the gravity/CYBE correspondence, as proposed in \cite{MY-LM}. 
This correspondence surely contains the relation between $r$-matrices and 
TsT transformations on coset spaces, but these are not all. 
Indeed, some examples presented in \cite{SUGRA-KMY} 
exhibit a curvature singularity in the middle of the bulk, 
but TsT transformations change only the asymptotic boundary behavior 
and would not lead to such a singularity. At the present moment, 
to what degree the gravity/CYBE correspondence can be extended is unknown. 
One of the motivations of this paper is to give a new example of the correspondence, 
which goes beyond the class of known cases and discuss possible further extensions.

\medskip

In this paper we consider type IIB superstrings on AdS$_5\times T^{1,1}$.
This geometry is realized by taking the near-horizon limit of 
a stack of $N$ D3-branes sitting at the tip of 
a conifold \cite{KW}. The internal manifold $T^{1,1}$ is a Sasaki-Einstein 
manifold with S$^2\times$S$^3$ topology and a $SU(2)\times SU(2)\times U(1)_{R}$ symmetry
(for details on the conifold see \cite{CD}, and for a review on aspects of AdS/CFT 
on this background see \cite{Herzog:2002ih}).  
At the present moment, the Green-Schwarz string action on this background has not been 
constructed. Thus, we will focus only on the bosonic sector. 

\medskip 

The usual description of $T^{1,1}$ as a coset is given by
\begin{equation}
\frac{SU(2) \times SU(2)}{U(1)}\,. 
\label{conventional}
\end{equation}
However, in this coset description one encounters a difficulty in applying the Yang-Baxter deformation to the usual coset decription of $T^{1,1}$, as we discuss now. Although (\ref{conventional}) describes the space topologically, the coset metric is not  the Sasaki-Einstein metric that the space admits\footnote{
This is well known and has been discussed in \cite{Castellani:1983tb,R}, 
where a general method for obtaining Einstein metrics on cosets was developed. 
However, this method does not seem suited for the study of the deformations 
we discuss here--see Appendix~\ref{app:A} for a discussion on this issue.}, and the one which is required as a proper  string background.
Since the class of deformations we are interested in are based on the coset 
description of the undeformed metric, before discussing deformations of $T^{1,1}$
we must develop a coset description that automatically leads to the Sasaki-Einstein metric. 

\medskip 

Our proposal is to describe $T^{1,1}$ as the bosonic part of the \textit{supercoset}\footnote{
Although the groups appearing below are bosonic, we refer to this as a supercoset 
due to a particular grading which is chosen. This will be discussed in detail in the main text.}:
\begin{align}
T^{1,1} = \frac{SU(2)\times SU(2)\times U(1)_R}{U(1)_1\times U(1)_2}\,.
\label{T11-coset intro}
\end{align}
As we shall show, it is possible to choose an embedding of the $U(1)$'s in the denominator that directly leads to the standard Sasaki-Einstein metric on $T^{1,1}$. In addition to leading to the correct undeformed metric, the description (\ref{T11-coset intro}) has the advantage that one can easily describe the general (three-parameter) deformation of this space, as a consequence of the explicit appearance of the $U(1)_{R}$ symmetry
in the numerator. This is rather natural given that $U(1)_{R}$ is part of the full global symmetry, 
and the grading of the matrices is rather natural from the point of view of the $\mathcal N=1$ superconformal 
symmetry of the dual gauge theory. It would be interesting to study whether this supercoset is relevant to the 
construction of the Green-Schwarz action on this background. 
The first step in this direction would be to find an appropriate supersymmetric extension by including fermions. 
However, the simplest extension (discussed below) will not contain 32 fermionic degrees of freedom and it may be difficult 
to construct the full Green-Schwarz action, as is usually the case in theories with reduced supersymmetry.

\medskip 

Next, we consider a family of three-parameter deformations of $T^{1,1}$ as 
Yang-Baxter sigma models with classical $r$-matrices satisfying the CYBE.  
This is analogous to the three-parameter real $\gamma$-deformations of S$^5$ as discussed in \cite{Frolov}. 
The resulting metric and NS-NS two form exactly agree with the ones obtained via TsT transformations in 
\cite{CO} and it contains the Lunin-Maldacena background \cite{LM} as a special case. 
This agreement  indirectly supports that the proposed supercoset description 
is the appropriate description of bosonic strings on AdS$_{5}\times T^{1,1}$\,. 

\medskip 

It is worth making a comment regarding the issue of integrability for $T^{1,1}$. 
Although it is generally believed that an integrability structure is present in some sectors, 
it was argued in \cite{BZ} that integrability for the full theory is lost due to the appearance 
of chaos in a certain subsector. Assuming that this conclusion is correct, 
our result indicates that the Yang-Baxter sigma model approach is applicable even for  non-integrable cosets.
This observation suggests that the gravity/CYBE correspondence  can be extended 
beyond integrable cases; integrability is not essential for the correspondence 
and it is just the tip of an iceberg. 

\medskip 

This paper is organized as follows. 
Section \ref{Sec:2} considers a coset construction of $T^{1,1}$\,. 
A supercoset description is proposed. 
In Section \ref{Sec:3}, we consider a family of deformations of $T^{1,1}$ as Yang-Baxter sigma model approach. 
We first give a short introduction to the Yang-Baxter sigma model approach. 
Then, the one-parameter deformation of $T^{1,1}$ is presented. 
Finally, three-parameter deformations are considered. 
Section \ref{Sec:4} is devoted to conclusion and discussion. 
Appendix \ref{app:A} reviews an alternative way to derive the $T^{1,1}$ metric.  
In Appendix \ref{app:B}, we give the detailed derivation of three-parameter deformation of $T^{1,1}$\,.

\section{A coset construction of \texorpdfstring{$T^{1,1}$}{T11}} 
\label{Sec:2}

In this section, we consider a coset construction of the $T^{1,1}$ metric\,. 
Instead of the conventional coset (\ref{conventional}), we describe the supercoset 
(\ref{T11-coset intro})\footnote{P.M.C. would like to thank Martin Ro\v cek for discussions on a related issue that inspired this construction.}.

\subsection{The \texorpdfstring{$T^{1,1}$}{T11} metric}

The internal manifold $T^{1,1}$ is a five-dimensional Sasaki-Einstein manifold with global isometry $SU(2)\times SU(2)\times U(1)_R$\,. 
The standard metric on $T^{1,1}$ is given by \cite{CD} 
\begin{align}
ds^2_{T^{1,1}} &=\frac{1}{6}(d\theta_1^2+\sin^2\theta_1d\phi_1^2)
+ \frac{1}{6}(d\theta_2^2+\sin^2\theta_2d\phi_2^2) \nonumber \\ 
& \qquad +\frac{1}{9}(d\psi+\cos\theta_1d\phi_1+\cos\theta_2d\phi_2)^2\,.   
\label{T11met} 
\end{align} 
This geometry may be regarded as a $U(1)$-fibration over S$^2\times$S$^2$\,. 
Here $0\leq \theta_i<\pi$ and $0\leq \phi_i<2\pi$
($i=1,2$) are the angle variables on two two-spheres. 
Then $0\leq \psi<4\pi$ is the coordinate along the $U(1)$-fiber.

\subsection{A supercoset representation of \texorpdfstring{$T^{1,1}$}{T11}}

As we have discussed, the coset representation (\ref{conventional}) 
does not lead to the metric (\ref{T11met}). Consider instead the following coset: 
\begin{align}
T^{1,1} = \frac{SU(2)\times SU(2)\times U(1)_R}{U(1)_1\times U(1)_2}\,. 
\label{T11-coset}
\end{align}
The generators of the two $\alg{su}(2)$'s and the $\alg{u}(1)_R$ 
in the numerator of (\ref{T11-coset}) are denoted by $K_i,\,L_i$ ($i=1,2,3$) and $M$\,, respectively.   
Rather than $5\times 5$ bosonic matrices, we choose a fundamental 
representation in terms of $(4|1)\times (4|1)$ {\it supermatrices}, $i.e.$,
\begin{align}
K_i=-\frac{i}{2}\left(
\begin{array}{cc|c} 
\sigma_i & 0 \,&\, 0 \\ 
0 & 0 \,&\, 0  \\ \hline  
0 & 0 \,&\, 0  
\end{array}\right)\,,
\qquad 
L_i=-\frac{i}{2}\left(
\begin{array}{cc|c} 
0 & 0 \,&\, 0 \\ 
0 & \sigma_i \,&\, 0  \\  \hline 
0 & 0 \,&\, 0  
\end{array}\right)\,,
\qquad 
M=-\frac{i}{2}
\left(
\begin{array}{cc|c} 
0 & 0 \,&\, 0 \\ 
0 & 0 \,&\, 0  \\ \hline  
0 & 0 \,&\, 1  
\end{array}\right)\,.
\label{sup-mat}
\end{align}
Here $\sigma_i$ ($i=1,2,3$) are the standard Pauli matrices,  
\begin{align}
\sigma_1=\begin{pmatrix}0 \,&\, 1 \\ 1 \,&\, 0\end{pmatrix}\,, \qquad 
\sigma_2=\begin{pmatrix}0 & -i \\ i & 0\end{pmatrix}\,, \qquad 
\sigma_3=\begin{pmatrix}1 & 0 \\ 0 & -1\end{pmatrix}\,. 
\end{align}
As we shall discuss below, the appearance of supermatrices-- 
rather than bosonic matrices--is in fact natural from the 
perspective of the full AdS$_{5}\times T^{1,1}$ coset space.

\medskip 

It is easy to see that the generators satisfy the following relations:  
\begin{align}
&[K_a,K_b]=\ep_{ab}{}^cK_c\,, \qquad [L_a,L_b]=\ep_{ab}{}^cL_c \,, \nonumber \\ 
&\str(K_aK_b)=\str(L_aL_b)=-\frac{1}{2} \de_{ab}\,, \qquad 
\str(MM)=\frac{1}{4}\,.   \nonumber 
\end{align}
Here the structure constant is normalized as $\ep_{123}=+1$
and the $\alg{su}(2)$ indices are raised and lowered by the Killing form $\de_{ab}$\,.  As usual, the supertrace of a supermatrix is defined as  
\begin{align}
\str\left(\begin{array}{c|c} 
A & B  \\ \hline 
C & D  
\end{array}\right) \equiv \Tr(A)-\Tr(D)\,, 
\end{align}
where $A,D$ are bosonic block matrices and $B,C$ are fermionic blocks. We denote the generators of the   
two $\alg{u}(1)$'s in the denominator of \eqref{T11-coset} by $T_{1,2}$ and we choose to embed them into the numerator by 
\begin{align}
T_1=K_3+L_3\,, \qquad T_2=K_3-L_3+4M\,.
\end{align}
Note that $T_{1}$ denotes the $U(1)$ in the usual description (\ref{conventional}). The final coset metric depends on the embedding of $T_{2}$ in the numerator, and we have chosen it such to obtain the Sasaki-Einstein metric (\ref{T11met}). 

\subsection{The \texorpdfstring{$T^{1,1}$}{T11} metric from a supercoset}

Let us first show that the supercoset (\ref{T11-coset}) indeed leads to the  metric (\ref{T11met}).  

\medskip 

It is convenient to introduce the orthogonal basis of the quotient vector space as follows: 
\begin{align}
\frac{\alg{su}(2)\oplus \alg{su}(2)\oplus \alg{u}(1)_R}{\alg{u}(1)_1\oplus \alg{u}(1)_2}
={\rm span}_{\mathbb{R}}\{K_1,K_2,L_1,L_2,H\}\,.
\label{coset-vs}
\end{align}
Here the diagonal element $H$ is defined as  
\begin{align}
H \equiv K_3-L_3+M\,. 
\end{align}
With this basis, one may introduce a group element parametrized by  
\begin{align}
g=\exp\bigl(\phi_1K_3+\phi_2L_3+2\psi M \bigr)
\exp\bigl(\theta_1K_2+(\theta_2+\pi)L_2\bigr)\,. 
\label{g-param} 
\end{align}
Then the left-invariant one-form  
\begin{align}
A \equiv g^{-1}d g \label{MC}
\end{align}
can be written in terms of the coordinates $\psi$, $\theta_i$ and $\phi_i$ ($i=1,2$)\,. 

\medskip 

The coset metric is given by the simple expression,
\begin{align} 
ds^2_{T^{1,1}}=-\frac{1}{3}\,\str\left[AP(A)\right]\,,  
\label{metric coset projector}
\end{align}
where $P$ is a projector to the coset space \eqref{coset-vs} and 
the associated projected current reads
\begin{align}
P(A)&=A+T_1\str[T_1A] -\frac{1}{3}T_2\str[T_2A]   \el \\
&= -\sin\theta_1d\phi_1\, K_1+d\theta_1\, K_2+\sin\theta_2d\phi_2\, L_1+d\theta_2\, L_2 
\el \\
&\qquad 
+\frac{2}{3}\bigl(d\psi+\cos\theta_1d\phi_1+\cos\theta_2d\phi_2\bigr)\, H\,.  
\label{proj}
\end{align} 
From this expression, it is direct to see that (\ref{metric coset projector}) leads to the metric (\ref{T11met}).

\subsection{What is the origin of the supercoset?}

Before discussing deformations of this space, it is worth discussing the origin of the supermatrix representations in \eqref{sup-mat}. A possible explanation is the following. It is believed that string theory on AdS$_5\times T^{1,1}$ 
is dual to an ${\cal N}=1$ superconformal field theory in four dimensions \cite{KW}.  
The ${\cal N}=1$ superconformal group is composed of the 
conformal group $SU(2,2)$\,, two sets of four real fermionic generators
$\overline{F}{}^A\,, F_A$, and the $U(1)_R$ symmetry. 
These generators can be organized into the  supermatrix, 
\begin{align}
\left(
\begin{array}{c|c} 
SU(2,2) & \overline{F}{}^A \\ \hline  
F_A  & U(1)_R  
\end{array}
\right)\,. 
\label{superconformal group}
\end{align}
Note that this supermatrix describes only the superconformal group $PSU(2,2|1)$, 
and does not contain the $SU(2)\times SU(2)$ flavor symmetry, unlike the case of $PSU(2,2|4)$ which includes the full flavor symmetry. 

\medskip 

Thus, to include flavor symmetry 
it is necessary to consider an embedding of $SU(2)\times SU(2)\times U(1)_{R}$ into a bigger supermatrix. 
A natural candidate is the following $(8|1)\times (8|1)$ supermatrix:  
\begin{align}
\left(
\begin{array}{cc|c} 
SU(2) & 0 & 0 \\ 
 0 & SU(2) & 0  \\ \hline  
 0 & 0 & U(1)_R  
\end{array}
\right)
\quad \hookrightarrow  \quad 
\left(
\begin{array}{ccc|c} 
SU(2,2) & 0 & 0 & \overline{F}{}^A \\ 
0 & SU(2) & 0 & 0 \\ 
0 & 0 & SU(2) & 0  \\ \hline  
F_A & 0 & 0 & U(1)_R  
\end{array}
\right)\,.  
\label{supermat} 
\end{align}
Here $PSU(2,2|1)$ is located at the four corners of (\ref{supermat}). 
Thus, the bosonic sector of the supercoset 
\begin{align}
\frac{PSU(2,2|1)\times SU(2)\times SU(2)}{SO(1,4)\times U(1)\times U(1)}
\end{align}
describes the bosonic sector of type IIB strings on $AdS_{5}\times T^{1,1}$. 
This is indeed a rather natural description of the full 
$PSU(2,2|1)\times SU(2)\times SU(2)$ symmetry group  and it may explain 
the origin of the supermatrix representation \eqref{sup-mat}\footnote{
It would be interesting to study whether turning on the fermions in this supercoset 
sigma model  is relevant for the construction of the Green-Schwarz action 
in this background, but we do not discuss this here.}.
As we shall discuss in Section~\ref{Sec:3}, the Yang-Baxter deformation of this supercoset 
leads to a family of deformations of the metric and NS-NS two-form that exactly agree 
with the ones obtained in \cite{CO}. The Lunin-Maldacena deformation \cite{LM} is 
contained as a special case. We consider this fact as further support for the supermatrix description.
It would be quite interesting to find further support for this interpretation from other points of view.

\section{Deformations of \texorpdfstring{$T^{1,1}$}{T11} as Yang-Baxter sigma models} 
\label{Sec:3}

Thus far, we have presented a supercoset construction of the Sasaki-Einstein metric on $T^{1,1}$. 
In this section we use this description to study Yang-Baxter deformations.

\medskip

By specifying classical $r$-matrices, we first discuss a one-parameter deformation 
in subsection \ref{Subsec:3.2} and then a three-parameter deformation in subsection 
\ref{Three-parameter deformation}.

\subsection{The action of Yang-Baxter sigma models on \texorpdfstring{$T^{1,1}$}{T11}}  

An interesting class of deformations of nonlinear sigma models is given by 
Yang-Baxter sigma models \cite{Klimcik,DMV}. 
The original procedure depends on the classical $r$-matrix of Drinfeld-Jimbo type, which 
satisfies the modified CYBE (mCYBE). 
However, this approach is not applicable to partial deformations\footnote{This point is explained as follows. The mCYBE for a Lie algebra $\alg{g}$ takes the form,
\begin{align}
[R(x),R(y)] - R([R(x),y]+[x,R(y)]) = c^2 [x,y] \qquad \text{for}\qquad {}^\forall x,y\in \alg{g} 
\nonumber  
\end{align} 
with a parameter $c$\,. To consider a partial deformation of a certain subalgebra $\alg{h}\subset \alg{g}$\,, 
the $R$-operator needs to satisfy $R(\alg{h})\subset \alg{h}$ and 
$R(\alg{m})=0$, where $\alg{m}$ is defined as $\alg{g}=\alg{h}\oplus\alg{m}$\,. 
From the mCYBE, this demands that the following two conditions be satisfied; 
(i) $c=0$ or $\alg{m}$ is abelian, and 
(ii) $R([R(x),y])=-c^2[x,y]$ for any $x \in \alg{h}$ and $y \in \alg{m}$\,. 
Note that, when $x \in \alg{m}$ and $y \in \alg{h}$\,, it also requires the condition (ii)
since the mCYBE is invariant by exchanging $x$ and $y$\,. 
Obviously, the $R$-operator of Drinfeld-Jimbo type does not satisfy these conditions. 
For $c\neq0$\,, we have found no solution. 
}. 
Since here we are interested  in deformations of the internal manifold $T^{1,1}$ only, 
we  apply the formalism of Yang-Baxter sigma models based on the CYBE \cite{KMY-Jor} instead. 

\medskip 

Our original motivation is to study type IIB superstrings on AdS$_5\times T^{1,1}$, and its deformations.  
However, since the Green-Schwarz action for these backgrounds 
have not been constructed, we restrict ourselves to the bosonic sector. 
For simplicity, we consider deformations of the internal manifold $T^{1,1}$ 
only (the AdS$_{5}$ part is untouched) and therefore we focus on this part of the action.

\medskip 

The action is given by 
\begin{align}
S=\frac{1}{3}(\ga^{\al\be}-\ep^{\al\be}) 
\int^\infty_{-\infty}\!\! d\tau \int^{2\pi}_{0}\!\! d\sigma\,   
\str\Bigl(A_\al P\circ \frac{1}{1-2\eta R_g\circ P}A_\be \Bigr)\,, 
\label{action}
\end{align}
where the flat metric $\ga^{\al\be}$ and the anti-symmetric tensor 
$\ep^{\al\be}$ on the string world-sheet
are normalized as $\ga^{\al\be}={\rm diag}(-1,1)$
and $\ep^{\tau\sigma}=1$\,. The projector $P$ to the coset space is given in \eqref{proj}\,. 
Here $\eta$ is a  parameter that measures deformations from $T^{1,1}$\,. 
In the $\eta\to 0$ limit, 
the action \eqref{action} reduces to the undeformed $T^{1,1}$, as shown in Section~\ref{Sec:2}. 

\medskip 

The left-invariant one-form is defined as  usual by
\begin{align}
A_\al\equiv g^{-1} \partial_\al g\,, \qquad 
g\in SU(2)\times SU(2)\times U(1)_R\,.  
\end{align}
The group element $g$ is parameterized as \eqref{g-param}. Note that the supertrace appears 
in the action (\ref{action}), even though all the fermions are set to zero in the present case.

\medskip 

The most important ingredient in (\ref{action}) is a linear $R$-operator. 
The symbol $R_g$ denotes a dressed $R$-operator, given by the adjoint operation of the group, as:
\begin{align}
R_g(X)\equiv g^{-1}R(gXg^{-1})g\,.
\label{Rg}
\end{align}
It is easy to see that if $R$ satisfies the CYBE, so does $R_{g}\,$. This $R$-operator is related to the tensorial notation of a classical $r$-matrix through 
\begin{align}
&R(X)=\str_2[r(1\otimes X)]=\sum_i \bigl(a_i\str(b_iX)-b_i\str(a_iX)\bigr) 
\label{linearR} \\
&\text{with}\quad r=\sum_i a_i\wedge b_i\equiv \sum_i (a_i\otimes b_i-b_i\otimes a_i)\,. \el 
\end{align}
In our case, $a_i$ and $b_i$ are generators in 
$\alg{su}(2)\oplus\alg{su}(2)\oplus\alg{u}(1)_R$\,. 

\subsection{One-parameter deformation} 
\label{Subsec:3.2}

We now consider examples of $r$-matrices describing deformations of $T^{1,1}$\,. 

\medskip 

Let us begin with the simplest example. This is provided by the abelian $r$-matrix,  
\begin{align}
r^{(\mu)}_{\rm Abe} = \mu K_3\wedge L_3\,,  
\label{r-1para}
\end{align}
with deformation parameter $\mu$\,. 
Here $K_3$ and $L_3$ are the Cartan generators 
of two $\alg{su}(2)$'s, respectively.  
The fundamental representation is given in \eqref{sup-mat}\,. 

\medskip 

Then the Lagrangian \eqref{action} is given by 
\begin{align}
&L =\frac{1}{3} (\ga^{\al\be}-\ep^{\al\be})\str\left[A_\alpha P(J_\be) \right]  \\
&\text{with}\qquad 
J_\be\equiv \frac{1}{1-2\bigl[R_{\rm Abe}^{(\mu)}\bigr]_g\circ P}A_\be\,,
\end{align}
where we have set the scaling factor $\eta=1$ in the deformed action\footnote{In fact, $\eta$ can be  
absorbed into the normalization of the $r$-matrices satisfying the CYBE.}. 
The operator $R_{\rm Abe}^{(\mu)}$ associated with \eqref{r-1para} 
is determined by the relation \eqref{linearR}\,. 
It is convenient to separate the Lagrangian into the two parts $L=L_G+L_B$\,, 
where $L_G$ is the metric part and $L_B$ is the coupling to the NS-NS two-form:  
\begin{eqnarray}
L_G &\equiv& -\frac{1}{3} [\str(A_{\tau}P(J_{\tau}))-\str(A_{\sigma}P(J_{\sigma}))] \,, \el \\ 
L_B &\equiv& -\frac{1}{3} [\str(A_{\tau}P(J_{\sigma}))-\str(A_{\sigma}P(J_{\tau}))] \,.  
\label{LGLB}
\end{eqnarray}

\medskip 

To evaluate the Lagrangian explicitly, it is sufficient to compute 
the projected current $P(J_\al)$ rather than $J_\al$ itself.  
Hence the computation is reduced to solving the following set of equations,  
\begin{align}
\left(1-2 P\circ \bigl[R_{\rm Abe}^{(\mu)}\bigr]_g\right)P(J_\al)=P(A_\al)\,. 
\label{rel}
\end{align}
Plugging the expression for $P(A_\al)$ given in \eqref{proj} into (\ref{rel}), one can solve for  
the deformed projected current, finding 
\begin{align}
P(J_\al)=j_\al^1\,K_1+j_\al^2\,K_2 
+j_\al^3\,L_1 +j_\al^4\,L_2+j_\al^5\,H\,, 
\end{align}
with the coefficients 
\begin{align}
j_\al^1&= \frac{G(6\mu)}{6}\sin\theta_1
\bigl[ (-6+4 \mu \cos \theta_1\cos\theta_2) \partial_\al\phi_1 
+ \mu(5-\cos 2\theta_2)\partial_\al \phi_2
\el \\
&\hspace{30mm}
+4 \mu (\cos\theta_2 +\mu\cos\theta_1 \sin ^2\theta_2 )\partial_\al \psi
\bigr] \,,\el \\ 
j_\al^2&=\partial_\al\theta_1 \,,\el \\ 
j_\al^3&= \frac{G(6\mu)}{6}\sin\theta_2
\bigl[ (6+4 \mu \cos \theta_1\cos\theta_2) \partial_\al\phi_2 
+\mu(5-\cos 2\theta_1)\partial_\al \phi_1
\el \\
&\hspace{30mm}
+4 \mu (\cos\theta_1 -\mu\cos\theta_2 \sin ^2\theta_1 )\partial_\al \psi
\bigr] \,,\el \\ 
j_\al^4&=\partial_\al\theta_2 \,,\el \\ 
j_\al^5&=\frac{2G(6\mu)}{3}
\bigl[ (\cos\theta_1 + \mu \sin ^2\theta_1 \cos\theta_2) \partial_\al\phi_1 
+ (\cos\theta_2 - \mu \sin ^2\theta_2 \cos\theta_1)\partial_\al \phi_2
\el \\
&\hspace{22mm}
+(1+\mu ^2 \sin ^2\theta_1 \sin ^2\theta_2)\partial_\al \psi\bigr]  \,,  
\end{align}
where the scalar function $G(x)$ is defined as 
\begin{align}
G(x)^{-1} \equiv 1+x^2 \left(\frac{\sin^2\theta_1\sin^2\theta_2}{36}
+\frac{\cos^2\theta_1\sin^2\theta_2+\cos^2\theta_2\sin^2\theta_1}{54}\right)\,. 
\end{align}
The resulting $L_G$ and $L_B$ are given by 
\begin{align}
L_G&=-\ga^{\al\be}G(\hat\ga) 
\Bigl[
\frac{1}{6}\sum_{i=1,2}\left(G(\hat\ga)^{-1} \partial_\al \theta_i\partial_\be \theta_i
+\sin^2\theta_i\partial_\al \phi_i \partial_\be \phi_i\right) 
+\hat\ga^2\frac{\sin ^2\theta_1\sin ^2\theta_2 }{324} \partial_\al\psi\partial_\be\psi 
\el \\
&\hspace{15mm} 
+\frac{1}{9}(\partial_\al \psi+\cos\theta_1\partial_\al \phi_1+\cos\theta_2\partial_\al \phi_2)
(\partial_\be \psi+\cos\theta_1\partial_\be \phi_1+\cos\theta_2\partial_\be \phi_2)
\Bigr]\,, \\
L_B&=2\ep^{\al\be} \hat\ga G(\hat\ga) \Bigl[
\frac{\cos\theta_2 \sin^2\theta_1}{54} \partial_\al\phi_1\partial_\be\psi  
-\frac{\cos\theta_1 \sin^2\theta_2}{54}\partial_\al\phi_2\partial_\be\psi  
\el \\
&\hspace{15mm}
+\Bigl(\frac{\sin^2\theta_1\sin^2\theta_2}{36}
+\frac{\cos^2\theta_1\sin^2\theta_2+\cos^2\theta_2\sin^2\theta_1}{54}\Bigr)
\partial_\al\phi_1\partial_\be\phi_2 \Bigr] \,,  
\end{align} 
where the new quantity $\hat{\gamma}$ is defined as  
\begin{align}
\hat\ga\equiv -6 \mu\,. 
\end{align}

\medskip 

Thus, the deformed metric and NS-NS two-form are given by  
\begin{align}
ds^2&=G(\hat\ga) 
\Bigl[
\frac{1}{6}\sum_{i=1,2}\left(G(\hat\ga)^{-1} d\theta_i^2 
+\sin^2\theta_i d\phi_i^2 \right) 
+\hat\ga^2\frac{\sin ^2\theta_1\sin ^2\theta_2 }{324} d\psi^2 
\el \\
&\hspace{15mm} 
+\frac{1}{9}(d\psi+\cos\theta_1d\phi_1+\cos\theta_2d\phi_2)^2 
\Bigr]\,, \\
B_2&=\hat\ga G(\hat\ga) \Bigl[
\frac{\cos\theta_2 \sin^2\theta_1}{54} d\phi_1\wedge d\psi  
-\frac{\cos\theta_1 \sin^2\theta_2}{54} d\phi_2\wedge d\psi  
\el \\
&\hspace{15mm}
+\Bigl(\frac{\sin^2\theta_1\sin^2\theta_2}{36}
+\frac{\cos^2\theta_1\sin^2\theta_2+\cos^2\theta_2\sin^2\theta_1}{54}\Bigr)
d\phi_1\wedge d\phi_2 \Bigr] \,.   
\end{align} 
These expressions agree exactly with the one-parameter $\ga$-deformed backgrounds 
presented by Lunin and Maldacena \cite{LM}\,.
Thus, the abelian $r$-matrix \eqref{r-1para} is the algebraic origin 
of the $\ga$-deformation of AdS$_5\times T^{1,1}$\,.

\subsection{Three-parameter deformation} 
\label{Three-parameter deformation} 

It is straightforward to generalize the one-parameter case to the three-parameter case. 
Since there are three Cartan generators $L_3, K_3$ and $M$\,, 
the most generic form  for the abelian $r$-matrix is given by  
\begin{align}
r^{(\mu_1,\mu_2,\mu_3)}_{\rm Abe} 
=\mu_1 L_3\wedge M+\mu_2 M\wedge K_3 +\mu_3 K_3\wedge L_3\,, 
\label{abe-3para}
\end{align}
with three deformation parameters $\mu_1, \mu_2$ and $\mu_3$\,. 
Note that the explicit appearance of the $U(1)_{R}$ symmetry--generated by $M$--
in the supercoset (\ref{T11-coset}) 
allows us to consider this three-parameter deformation.

\medskip 

The computation is completely parallel to the one-parameter case.  
Thus, we do not repeat it here but simply give the final result. 
For details, see Appendix \ref{app:3para}\,.

\medskip 

With parameter identifications\footnote{
Here we also normalize the scaling factor in \eqref{action} as $\eta=1$.} 
\begin{align}
3 \mu_1= \hat\ga_1 \,, \qquad 
3 \mu_2= \hat\ga_2\,, \qquad 
-6 \mu_3= \hat\ga_3\,,  
\end{align}
we obtain the following deformed metric and NS-NS two-form:
\begin{align}
ds^2 &= G(\hat\ga_1,\hat\ga_2,\hat\ga_3)\Bigl[
\frac{1}{6}\sum_{i=1,2}(G(\hat\ga_1,\hat\ga_2,\hat\ga_3)^{-1}d\theta_i^2+\sin^2\theta_i d\phi_i^2) \el \\
&\quad 
+\frac{1}{9}(d\psi+\cos\theta_1d\phi_1+\cos\theta_2d\phi_2)^2 
+\frac{\sin^2\theta_1\sin^2\theta_2}{324}(\hat\ga_3 d\psi+\hat\ga_1 d\phi_1+\hat\ga_2 d\phi_2)^2 \Bigr]\,, 
\label{3para-G}
\\
B_2 &=G(\hat\ga_1,\hat\ga_2,\hat\ga_3)\Bigl[
\Bigl\{\hat\ga_3 \Bigl(\frac{\sin^2\theta_1\sin^2\theta_2}{36}
+\frac{\cos^2\theta_1\sin^2\theta_2+\cos^2\theta_2\sin^2\theta_1}{54}\Bigr)\el \\
&\hspace{33mm}
-\hat\ga_2 \frac{\cos\theta_2\sin^2\theta_1}{54}
-\hat\ga_1 \frac{\cos\theta_1\sin^2\theta_2}{54} 
\Bigr\}d\phi_1\wedge d\phi_2 \el\\
&\quad 
+\frac{(\hat\ga_3\cos\theta_2-\hat\ga_2)\sin^2\theta_1}{54}d\phi_1\wedge d\psi 
-\frac{(\hat\ga_3\cos\theta_1-\hat\ga_1)\sin^2\theta_2}{54}d\phi_2\wedge d\psi 
\Bigr]\,, 
\label{3para-B}
\end{align}
where the scalar function is defined as  
\begin{align}
G(\hat\ga_1,\hat\ga_2,\hat\ga_3)^{-1}&\equiv1
+\hat\ga_3^2 \Bigl(\frac{\sin^2\theta_1\sin^2\theta_2}{36}
+\frac{\cos^2\theta_1\sin^2\theta_2+\cos^2\theta_2\sin^2\theta_1}{54}\Bigr)
+\hat\ga_2^2 \frac{\sin^2\theta_1}{54} \el \\
&\qquad 
+\hat\ga_1^2 \frac{\sin^2\theta_2}{54} 
-\hat\ga_2\hat\ga_3 \frac{\sin^2\theta_1\cos\theta_2}{27} 
-\hat\ga_3\hat\ga_1 \frac{\sin^2\theta_2\cos\theta_1}{27} \,. 
\label{3para-Gfun}
\end{align}
These expressions are rather complicated but  agree perfectly with the ones obtained in \cite{CO}\,. 
Thus, the abelian $r$-matrix \eqref{abe-3para} corresponds to the
three-parameter $\ga$-deformation. 
The previous one-parameter deformation is reproduced by simply setting 
$\hat\ga_1 =\hat\ga_2 =0$ and $\hat\ga_3 =\hat\ga $\,.

\medskip 

Finally, let us comment on the amount of supersymmetry remaining in the three-parameter deformation. 
Recall that  in the undeformed $T^{1,1}$ case there is an $\mathcal{N}$=1 superconformal symmetry. 
Without studying the Killing spinor equations, we can anticipate that no supersymmetry should remain 
for a generic value of the parameters. Note that in the classical $r$-matrix (\ref{abe-3para}), 
the generator $M$ is associated with the corresponding $U(1)$ R-symmetry, 
while $K_3$ and $L_3$ are associated to the non-R symmetry $SU(2)\times SU(2)$\,. 
In the Lunin-Maldacena case \cite{LM} with $\mu_3\neq 0$ and $\mu_1=\mu_2=0$\,, 
the $\mathcal{N}$=1 superconformal symmetry is preserved because the $U(1)$ R-symmetry 
is not affected by the TsT transformation. However, if either $\mu_1$ or $\mu_2$ is non-zero,
the solution is non-supersymmetric\footnote{Note that the background still seems to preserve 
the $U(1)$ R-symmetry. However, one should be careful with the periodicity of the angle variables and note that the Killing spinors cannot survive for generic values of $\mu_1$ and $\mu_2$\,. This is a global property and 
cannot be seen from a local quantity like the metric.}.

\section{Conclusion and discussion}
\label{Sec:4}

In this paper we have considered a family of deformations of $T^{1,1}$ 
as Yang-Baxter sigma models. 

\medskip 

We first provided a new coset description of $T^{1,1}$ 
which directly leads to the standard Sasaki-Einstein metric. 
This is necessary to study deformations of this space as Yang-Baxter sigma models. 
The coset description we presented is a rather natural description from the point of view of the $\mathcal N=1$ 
superconformal symmetry of the dual gauge theory. However, to the best of our knowledge this  description 
has not appeared in the literature.

\medskip 
 
Next, we considered three-parameter deformations of $T^{1,1}$ 
by using classical $r$-matrices satisfying the CYBE.
The resulting metric and NS-NS two-form perfectly agree with 
the ones obtained via TsT transformations \cite{LM,CO}. 

\medskip 

It was shown in \cite{MY-LM} that three-parameter real $\ga$-deformations AdS$_5\times$S$^5$ \cite{LM,Frolov}
are realized by the Yang-Baxter sigma model approach with abelian classical $r$-matrices. 
Thus, the results obtained here may be regarded as a generalization of the work \cite{MY-LM},  
giving further support for the gravity/CYBE correspondence. 
However, it should be stressed that there is a significant difference 
between S$^5$ and $T^{1,1}$\,. The former is represented by a symmetric coset 
and therefore corresponds to an \textit{integrable} nonlinear sigma model. In the case of $T^{1,1}$, however, this is not the case and the claim that 
it is not integrable was made in \cite{BZ}, by showing the appearance of chaos in a subsector of the theory. 
Assuming that this result is correct, the class of deformations considered here are not regarded as 
integrable deformations. However, this would lead to the stronger statement that 
the gravity/CYBE correspondence would hold independently 
of integrability and that it captures a much wider class of gravitational solutions.

\medskip 

Let us make a few comments on possible further generalizations. 
An interesting class of metrics on S$^{2}\times$S$^{3}$ is given by the well-known $Y^{p,q}$ metrics 
\cite{Gauntlett:2004yd}. However, since these have not been explicitly constructed as coset metrics, 
it would be difficult to consider deformations in this approach.
It would also be interesting to study additional coset spaces which may or may not be integrable, a possible candidate being the Lifshitz spacetime. The coset description was given in \cite{SYY}, 
and it has been argued to be non-integrable in \cite{GS}. Other important supercosets appear in descriptions of type IIA compactifications on AdS$_{4}$, 
such as ABJM theory \cite{Aharony:2008ug}. The supercoset description has been given in \cite{Arutyunov:2008if}.

\medskip

What is the general class of gravitational solutions included in the gravity/CYBE correspondence?
As we have discussed above, it has already been shown that the correspondence includes deformations which cannot be obtained by TsT transformations. The result obtained in this paper indicates that the integrability of the parent theory is not an essential feature. Thus, we see that the class of gravitational solutions captured by the correspondence is much wider than the examples that were first discovered.  What  the full moduli space of  gravity solutions captured by the gravity/CYBE correspondence is remains an open problem at the present moment.

\medskip

As we have seen, at this point there are various examples of coset supergravity backgrounds, integrable and non-integrable,  such that its Yang-Baxter deformations remain as supergravity solutions. The non-trivial question is whether this is the case for a generic coset supergravity background and a generic $r$-matrix. Although a counter-example has not been found so far, there is no proof that this is true in general. One possible approach to studying this would be to exploit kappa-symmetry.  Answering this question could lead to new insights into the structure of the moduli space of possible gravity solutions, and the action of classical $r$-matrices on this space. This issue deserves to be studied as a fundamental problem.

\subsection*{Acknowledgments}

We are very grateful to Gleb Arutyunov, Riccardo Borsato, Martin Ro\v cek and Stefan Vandoren 
for valuable comments and discussions. 
We also thank Stijn van Tongeren for correspondence. 
P.M.C. is supported by the Netherlands Organization for Scientific Research
(NWO) under the VICI grant 680-47-603. T.M.\ is supported by the Netherlands Organization for Scientific 
Research (NWO) under the VICI grant 680-47-602.  This work is also part of the ERC Advanced grant research programme 
No.~246974, ``Supersymmetry: a window to non-perturbative physics" 
and of the D-ITP consortium, a program of the NWO that is funded by the 
Dutch Ministry of Education, Culture and Science (OCW).

\appendix 

\section{\texorpdfstring{$T^{1,1}$}{T11} metric from the rescaling of vielbeins}
\label{app:A}

As we have discussed, the $(SU(2)\times SU(2))/U(1)$ coset description of $T^{1,1}$ does not lead to 
the Sasaki-Einstein metric (\ref{T11met}) that the space admits. 
This comes as no surprise, since it is well known that coset spaces are not typically Einstein spaces. 
However,  it was shown in \cite{Castellani:1983tb} that given a coset space $G/H$ 
it may be possible to rescale the vielbeins to obtain an Einstein space, 
without loosing the original symmetry of the coset space. 
This is in fact the case for $T^{1,1}$, as discussed in \cite{R}. 
Take the left-invariant current $A=g^{-1}dg$ with $g\in SU(2)\times SU(2)$
and rescale the coset space directions 
by three parameters $\alpha, \beta, \gamma$, as
\equ{
A_{resc.}=\alpha \sum_{i=1,2} A^{i}K_{i}+ \beta \sum_{i=1,2} A^{i}L_{i}
+\gamma \, A^{-}(L_{3}-K_{3})+A^{+}(L_{3}+K_{3})\,.
\label{rescaled current}
}
The term proportional to $A^{+}$ is the one projected out by the coset and is not rescaled. 
For $\alpha=\beta=\ga=1$, this current describes a natural metric on the coset space 
$(SU(2)\times SU(2))/U(1)$ but not the Sasaki-Einstein metric. 
However, for arbitrary values of the parameters one finds\footnote{
A more general metric is obtained by taking the general invariant two-form into account \cite{T11-general}.}
\equ{
ds^{2}= \alpha^{2} (d\theta_1^2 +\sin^2 \theta_1\, d \phi_1^2) 
+ \beta^{2} (d\theta_2^2 +\sin^2 \theta_2 \, d \phi_2^2 ) 
+\frac{\gamma^{2}}{2}\(d \psi +\cos \theta_1\, d \phi_1 +\cos \theta_2\, d\phi_2\)^2\,.
\label{metric parameters}
}
Imposing the Einstein condition on this metric one finds
\equ{
\alpha^{2} =\beta^{2}= \frac{1}{6} \,,\quad \gamma^{2} =\frac{2}{9}\,,
\label{CY parameters}
}
corresponding to (\ref{T11met}). Thus, a possible starting point to study deformations of the $T^{1,1}$ 
sigma model would be to study deformations of the sigma model defined by the rescaled current 
(\ref{rescaled current}). However,  since this approach is based on a rescaling of the \textit{current}, 
rather than the group elements $g$, is not clear how to implement the Yang-Baxter deformation 
(defined by the action of the group elements in (\ref{Rg})) in this formulation. 
Thus, one of the advantages of the supercoset description (\ref{T11-coset}) is that 
the Yang-Baxter deformation can be applied directly, as we have shown.  
Another advantage is that by making manifest the $U(1)_{R}$ symmetry, 
it is clear how to implement the three-parameter deformation discussed in 
Section~\ref{Three-parameter deformation}.

\medskip 

As a final comment, we would like to point out that a related issue arises 
in the description of the conifold as a classical K\"{a}hler quotient. 
It is well known that this can be realized as an $\mathcal N=(2,2)$ gauged linear sigma model (GLSM) 
for four chiral fields with charges $(1,1,-1,-1)$ under a $U(1)$ \cite{KW}. 
It is easy to see that the classical quotient metric is not the Calabi-Yau metric, $i.e.$, 
the metric of the base is not the Sasaki-Einstein metric (in fact, it coincides with the coset metric).
Again, this comes as no surprise since the classical quotient metric describes 
the UV behavior of the GLSM, while the Calabi-Yau metric describes 
the IR behavior, at the endpoint of the RG flow. 
It would be interesting to study whether it is possible to formulate the supercoset 
description of the conifold that we have given here 
in terms of a GLSM\footnote{We would like to thank Martin Ro\v cek for discussions on this.}.

\section{Derivation of three-parameter deformations  \label{app:3para}}
\label{app:B}

It would be useful to present here the detailed derivation of the 
three-parameter deformed metric \eqref{3para-G} and 
NS-NS two-form \eqref{3para-B}\,. 

\medskip 

The classical $r$-matrix is composed of 
three Cartan generators $L_3, K_3$ and $M$ as follows: 
\begin{align}
r^{(\mu_1,\mu_2,\mu_3)}_{\rm Abe} 
=\mu_1 L_3\wedge M+\mu_2 M\wedge K_3 +\mu_3 K_3\wedge L_3\,.   
\end{align}
Here $\mu_1$, $\mu_2$ and $\mu_3$ are deformation parameters. 
Then the associated linear R-operator is written in terms of $L_3, K_3$ and $M$ like 
\begin{align}
&R^{(\mu_1,\mu_2,\mu_3)}_{\rm Abe}(K_3)= \frac{1}{2}(\mu_3 L_3-\mu_2 M)\,, & 
&R^{(\mu_1,\mu_2,\mu_3)}_{\rm Abe}(L_3)= \frac{1}{2}(\mu_1 M -\mu_3 K_3)\,, \el \\  
&R^{(\mu_1,\mu_2,\mu_3)}_{\rm Abe}(M)= \frac{1}{4}(\mu_1 L_3 -\mu_2 K_3)\,, & 
&R^{(\mu_1,\mu_2,\mu_3)}_{\rm Abe}(\text{others})=0\,. 
\end{align}
These transformation laws are utilized to rewrite the Lagrangian  \eqref{LGLB}\,. 

\medskip 

First of all, let us evaluate the projected deformed current $P(J_\al)$\,. 
It can be done by solving the relation, 
\begin{align}
\left(1-2 P\circ \bigl[R_{\rm Abe}^{(\mu_1,\mu_2,\mu_3)}\bigr]_g\right)P(J_\al)=P(A_\al)\,. 
\end{align}
Plugging the expression of $P(A_\al)$ in \eqref{proj} with the above equation, 
the deformed projected current is obtained as 
\begin{align}
P(J_\al)=j_\al^1\,K_1+j_\al^2\,K_2 +j_\al^3\,L_1 +j_\al^4\,L_2+j_\al^5\,H\,, 
\end{align}
with the coefficients 
\begin{align}
j_\al^1&= \frac{1}{6}G(3\mu_1,3\mu_2,-6\mu_3)\sin\theta_1
\el \\
&\quad \times\bigl[ 
-\left(6+\mu_1 \sin^2\theta_2 (\mu_1+2\mu_3 \cos\theta_1)
-2  \cos\theta_1(\mu_2+2\mu_3 \cos\theta_2)\right)\partial_\al\phi_1 
\el\\
&\qquad 
+ \left(2 \cos\theta_2 (\mu_2+2\mu_3 \cos\theta_2)
-\sin ^2\theta_2 (\mu_1\mu_2 -6\mu_3+2\mu_2\mu_3 \cos\theta_1)\right)
\partial_\al \phi_2
\el \\
&\qquad 
+2 \left(\mu_3 \sin^2\theta_2 (\mu_1+2\mu_3 \cos\theta_1)
+\mu_2 +2 \mu_3 \cos\theta_2 \right) \partial_\al \psi
\bigr] \,,\el \\ 
j_\al^2&=\partial_\al\theta_1 \,,\el \\ 
j_\al^3&= 
\frac{1}{6}G(3\mu_1,3\mu_2,-6\mu_3)\sin\theta_2
\el \\
&\quad \times\bigl[
\left(6+ \mu_2\sin^2\theta_1(2\mu_3 \cos\theta_2+\mu_2)
+2 \eta\cos\theta_2 (2\mu_3 \cos\theta_1+\mu_1)\right)\partial_\al \phi_2
\el\\
&\qquad 
+\left(2 \cos\theta_1 (\mu_1+2\mu_3 \cos\theta_1)
+\sin^2\theta_1 (\mu_1\mu_2 +6\mu_3+2\mu_1\mu_3 \cos\theta_2)\right)\partial_\al\phi_1 
\el \\
&\qquad 
+2 \left(-\mu_3 \sin^2\theta_1 (2\mu_3 \cos\theta_2+\mu_2)
+\mu_1+2 \mu_3 \cos\theta_1\right)\partial_\al \psi
\bigr] \,,\el \\ 
j_\al^4&=\partial_\al\theta_2 \,,\el \\ 
j_\al^5&=
\frac{1}{3}G(3\mu_1,3\mu_2,-6\mu_3)
\el \\
&\quad \times\bigl[
\left(2 \cos\theta_1+\sin^2\theta_1\left(\mu_2
-\mu_1\mu_3 \sin^2\theta_2+2\mu_3 \cos\theta_2\right)\right)\partial_\al \phi_1
\el\\
&\qquad 
+\left(2 \cos\theta_2-\eta \sin^2\theta_2\left(\mu_1
+\mu_2\mu_3 \sin^2\theta_1+2\mu_3 \cos\theta_1\right)\right)\partial_\al \phi_2
\el \\
&\qquad 
+2\left(1+ \mu_3^2 \sin^2\theta_1\sin^2\theta_2\right) \partial_\al \psi
\bigr] \,. 
\end{align}
Here the scalar function $G(\hat\ga_1,\hat\ga_2,\hat\ga_3)$ is defined in \eqref{3para-Gfun}\,. 

\medskip 

As a result, $L_G$ and $L_B$ are given by, respectively,  
\begin{align}
L_G &= -\ga^{\al\be}G(\hat\ga_1,\hat\ga_2,\hat\ga_3)\Bigl[
\frac{1}{6}\sum_{i=1,2}(G(\hat\ga_1,\hat\ga_2,\hat\ga_3)^{-1}\partial_\al\theta_i\partial_\be\theta_i
+\sin^2\theta_i \partial_\al\phi_i \partial_\be\phi_i) \el \\
&\quad 
+\frac{1}{9}(\partial_\al \psi+\cos\theta_1\partial_\al \phi_1+\cos\theta_2\partial_\al \phi_2)
(\partial_\be \psi+\cos\theta_1\partial_\be \phi_1+\cos\theta_2\partial_\be \phi_2)  \el \\
&\quad 
+\frac{\sin^2\theta_1\sin^2\theta_2}{324}
(\hat\ga_3 \partial_\al\psi+\hat\ga_1 \partial_\al\phi_1+\hat\ga_2 \partial_\al\phi_2)
(\hat\ga_3 \partial_\be\psi+\hat\ga_1 \partial_\be\phi_1+\hat\ga_2 \partial_\be\phi_2) \Bigr]\,, \\
L_B &=2\ep^{\al\be}G(\hat\ga_1,\hat\ga_2,\hat\ga_3)\Bigl[
\Bigl\{\hat\ga_3 \Bigl(\frac{\sin^2\theta_1\sin^2\theta_2}{36}
+\frac{\cos^2\theta_1\sin^2\theta_2+\cos^2\theta_2\sin^2\theta_1}{54}\Bigr)\el \\
&\hspace{40mm}
-\hat\ga_2 \frac{\cos\theta_2\sin^2\theta_1}{54}
-\hat\ga_1 \frac{\cos\theta_1\sin^2\theta_2}{54} 
\Bigr\}\partial_\al\phi_1\partial_\be\phi_2 \el\\
&\quad 
+\frac{(\hat\ga_3\cos\theta_2-\hat\ga_2)\sin^2\theta_1}{54}
\partial_\al\phi_1 \partial_\be\psi 
-\frac{(\hat\ga_3\cos\theta_1-\hat\ga_1)\sin^2\theta_2}{54}
\partial_\al\phi_2 \partial_\be\psi 
\Bigr]\,, 
\end{align}
with the following parameter identifications:  
\begin{align}
3 \mu_1= \hat\ga_1 \,, \qquad 
3 \mu_2= \hat\ga_2\,, \qquad 
-6 \mu_3= \hat\ga_3\,.   
\end{align}
Thus, the resulting metric and NS-NS two-form turn out to be \eqref{3para-G}
and \eqref{3para-B}, respectively.

\end{document}